\documentstyle[times,pramana,epsf,floats]{ias}

\def\f{\phi} 
\def\m{\mu} 
\def\n{\nu} 
\def\p{\pi} 
\def\r{\rho} 
\def\s{\sigma} 
\def\t{\tau} 
\renewcommand\O{\Omega}
\newcommand\aO{\bar \Omega}

\def\cl{\mathcal L}

\def\del{\partial}
\def\tr{\mbox{\rm tr}}
\def\exp{\mbox{\rm exp}}

\def\ra{\rightarrow} 

\def\bc{\begin{center}}
\def\ec{\end{center}}
\def\be{\begin{equation}}
\def\ee{\end{equation}}
\def\bea{\begin{eqnarray*}}
\def\eea{\end{eqnarray*}}

\begin{document}
\mark{{Possible evidence of DCC at the SPS}{J I Kapusta and S M H Wong}} 
\title{Possible Evidence of Disoriented Chiral Condensates from
       the Anomaly in $\O$ and $\aO$ Abundances at the SPS}

\author{J I Kapusta$^a$ and S M H Wong$^b$\footnote{The speaker. S.W. thanks
        the organizers for their tireless efforts and a nice conference.}} 
\address{$^a$School of Physics and Astronomy, University of Minnesota, 
         Minneapolis, MN 55455, USA \\
         $^b$Department of Physics, Ohio State University, Columbus, 
         OH 43210, USA}


\abstract{
No conventional picture of nucleus-nucleus collisions has yet been 
able to explain the abundance of $\O$ and $\aO$ in
central collisions between Pb nuclei at 158 A GeV at the CERN SPS.  
We argue that such a deviation from predictions of statistical 
thermal models and numerical simulations is evidence that they are produced 
as topological defects in the form of skyrmions arising from the 
formation of disoriented chiral condensates. The estimated domain size
falls in the right range to be consistent with the so far non-observation 
of DCC from the distribution of neutral pions. 
} 

\maketitle

\section{Introduction}

Ever since their introduction \cite{an,jk,bkt}, disoriented 
chiral condensates (DCC) have been the subject of intense theoretical 
efforts and the goal of experimental search. No sign of their formation
has so far been found. One might be tempted to jump to the conclusion that 
DCC did not form in any of the heavy ion collision experiments conducted
to date. However, this is not the only conclusion that can be drawn from
these experiments. DCC were supposed to manifest themselves 
in the distribution of neutral to total pion ratio only if 
a number of large domains with different chiral orientation were formed
in the collisions. If the domains are small then it is impossible to
tell, at least not via the pion ratio, whether DCC have been formed.  
We shall argue that it is possible to observe DCC via strange 
hyperons and that experimental data already provided some clues. 

Our motivation originated from data of $\O$ and $\aO$ yields at the 
Super Proton Synchrotron (SPS). The measure of how well statistical models  
fit to particle ratios, $\chi^2$, reduces by an order of magnitude if $\O$ 
and $\aO$ are left out \cite{rl}. The slope of the hadron $m_T$ spectra follows 
a linear increase with mass, but $\O$ and $\aO$ deviate significantly from 
this \cite{san,wa97}. This seems to be an indication that the triply 
strange hyperons do not follow the same trend of the other hadrons. 
It has been difficult for numerical models to generate the same yield 
of the strange hyperons at the SPS without re-adjusting parameters 
\cite{urqmd}. In general there tend to be more hyperons than expected.   
One might wonder if the excess over the expected yield comes from all
values of $k_T$ or if this has an origin in only a particular range.  
Such information might provide further clues to the source of the 
additional hyperons. It has been shown that the triply strange hyperon 
yield is systematically higher than expected from model chemical analysis 
only at low $k_T$ \cite{tr} so this points to a source of low momentum
$\O$ and $\aO$. We wish to remind readers at this point that DCC 
are also a source of low energy pions. If DCC can produce hyperons,
then it is in their nature as condensates to yield only low energy ones. 
But how can DCC give rise to baryons?

\section{Producing Hyperons from DCC}

While one might be accustomed to classical chiral fields giving rise
to low energy pions from the numerous recent works on DCC, it has been
pointed out much earlier in the 60's that chiral fields could produce
baryons \cite{sk,w1,w2,w3}. The essential connection is the Skyrme model
which has the Lagrangian density 
\be {\cl}_S = \frac{f_\p^2}{4} \; \tr(\del_\m U\del^\m U^\dagger) 
   +\frac{1}{32g^2}  \; \tr[U^\dagger \del_\m U,U^\dagger \del_\n U]^2 
\ee 
where $U = \exp \{i \mbox{\boldmath $\t$} \cdot \f /f_\p\} 
      = ( \s + i \mbox{\boldmath $\t$} \cdot \mbox{\boldmath $\p$})/f_\p $. 
Here $f_\p$ is the pion decay constant and $g$ is essentially the
$\p$-$\r$-$\p$ coupling constant. The equation of motion of this Lagrangian 
permits solutions called skyrmions. Spherically symmetric solutions 
can be found in the form 
\be U = U_S = \exp \{i \mbox{\boldmath $\t$}\cdot \mbox{\boldmath $\hat r$}\; 
                     F(r) \}
\ee
which depends on one radial function $F(r)$. It is required that
$F(r)$ satisfies the boundary conditions $F(r \ra \infty) \ra 0$ and 
$F(r=0) = N \p$. 
$N$ is an integer called the winding number and has been identified as the 
baryon number \cite{sk,w1,w2,w3}. For larger values of $N$ it is not clear
if the solutions really describe nuclei, but for the case of $N=\pm 1$  
it represents a baryon and antibaryon, respectively. Therefore classical
chiral fields can generate composite fermions. Unlike the case of pions,
it is not automatic that the fields will generate baryons. A non-trivial
field topology must be formed before baryons can be produced.

Since the potential of the Skyrme Lagrangian is too complicated to permit
easy understanding of condensates formation, let us discuss instead the 
case of a very simple version of the linear sigma model 
\be {\cl} = \frac{1}{2} \del_\mu \Phi^\alpha \del^\mu \Phi_\alpha 
           -\frac{\lambda}{4} (\Phi^\alpha \Phi_\alpha -v^2 )^2 
\label{eq:lsm} 
\ee
where $\Phi_\alpha = (\sigma,\pi_1,\pi_2,\pi_3)$. The potential is
totally symmetric with respect to the chiral fields. The vacuum of this
theory is simply the surface of a four dimensional sphere or $S^3$. In 
order to form non-trivial topology, there must be domain formation where 
the chiral condensates in neighbouring domains should have sufficiently
different chiral orientations so that they cover $S^3$ and would not be 
able to evolve back to the origin. The probability per unit volume 
for this to happen in the linear sigma model has been worked out 
\cite{ks,sp,lee,kw,kw2} in terms of the correlation length $\xi$ to be 
$0.04 \xi^{-3}$. Because numerically flavour $SU(2)\simeq SU(3)$ 
we can apply this result to the hyperons \cite{ek,ek2,eh}.

\section{$\O$ and $\aO$ Production from DCC at the SPS} 

To examine the possibility of skyrmion formation from DCC we have
to look at several things. How many hyperons originate from 
DCC per central collision? What is the likely domain size? It must 
not be too large so that DCC would manifest themselves in the distribution
of the pion ratio, contradicting observational evidence. Finally, any 
hyperons thus formed should not be easily destroyed via collisions. 

At the SPS the WA97 collaboration is the only experiment with published 
data of the yield of $\O$ and $\aO$ \cite{cal}. They gave 
$\aO/\O  = 0.383 \pm 0.081$ and $\O+\aO  = 0.410 \pm 0.08$. 
in an interval of $\Delta y=1$ centering around $y=0$. Unfortunately, this 
is quite a narrow range. We would like the $\O$ data in a wider rapidity
range and possibly in the whole of $4\pi$. The NA49 collaboration measured
also the hyperons except the $\O$ and $\aO$ \cite{mar,gab}. Furthermore,
their $\Xi$ results have been extrapolated to $4\pi$ in \cite{bc} to be
$\Xi^-  = 7.5  \pm 1.0$ and $\Xi^- + \bar \Xi^+  = 8.2  \pm 1.1$. 
These latter are useful for us because WA97 also measured $\Xi$ in the
same rapidity window as their $\O$ measurement. They are
$\Xi^-  = 1.50 \pm 0.10$ and $\bar \Xi^+  = 0.37 \pm 0.06$. 
The combination of these results allow us to extrapolate the $\O$
results to $4\pi$ as follows, 
\be \left (\frac{\aO}{\O+\aO}\right )_{\mbox{\scriptsize WA97}} 
    \left (\frac{\O+\aO}{\Xi^- +\bar \Xi^+} \right )_{\mbox{\scriptsize 
                                                            WA97}} 
    \Big (\Xi^- +\bar \Xi^+ \Big )_{\mbox{\scriptsize NA49}} = 0.498 \;.
\ee
One can expect on the average half an $\aO$ per central collision at the SPS. 

Since $\aO$ are rather rare, let us assume all of them are from DCC and that 
there is an equal chance of forming any of the octet or decuplet baryons.
(One might expect that strange baryons with the more massive strange quark
content should have a smaller probability. However, such flavour symmetry
breaking would allow skyrmions to form in a smaller spatial region. This
leads to larger probability per unit volume, thus compensating for the 
reduction in probability due to the larger mass \cite{ks,kw,kw2}). This gives 
about 7 skyrmions or antiskyrmions per central collision on the average.  
With about 2000 hadrons produced at the SPS and a not too unreasonable
assumption of DCC forming at a matter density of around 10 times normal 
nuclear matter of 1.7 hadrons/fm$^3$, one can estimate from the probability
per unit volume discussed in the previous section that the correlation
length or DCC domain size is $\xi \sim 2$ fm. This is too small a size for 
observation via pions and is consistent with theoretical expectation
based on the system evolving in time in thermal equilibrium \cite{rw}
or a slow time evolution of the effective potential back to the vacuum
form in the annealing scenario \cite{gm}. These scenarios gave 
$\xi \sim 1.5$ fm and $3-4$ fm, respectively. Note that the chance for
observing DCC through this mechanism of forming hyperons is completely 
orthogonal to that of forming low energy pions. The latter requires large
domains whereas the former needs small ones.

Now we have to try to answer the question of whether hyperons coming
from DCC can be easily destroyed or not. If they cannot survive the
onslaught of the hadronic medium, any signal of DCC would be washed out. 
To check this we write down the rate for destroying $\O$ or $\aO$
via the most numerous and energetic of hadrons, namely $\pi$ and $K$,
\be \frac{\partial \r_\O}{dt} = 
   -\langle \s_{_{  K + \Xi \rightarrow \p + \O}} v_{\p\O} \rangle R_5 
            \r_\p^\infty \r_{_\O} 
   -\langle \s_{_{ \p + \Xi \rightarrow  K + \O}} v_{ K\O} \rangle R_1 
            \r_K^\infty  \r_{_\O} \;. 
\label{eq:rd}
\ee
The notations are the same as those in \cite{kmr}. This assumes thermal
equilibrium of everything else except the $\O$ and universal interaction matrix
elements so that the difference due to flavours are completely in the phase
space. Using results contained in \cite{kmr} one can work out the destruction 
time-scales. For example at $T=200$ MeV, the time-scale for changing the 
$\O$ into something else by collision with pions is $\tau_{\p} \sim 257$ fm 
and that by $K$ is $\tau_K \sim 152$ fm. These are rather long times. 
One can conclude that there is a very good chance that hyperons from DCC can 
show up in the detectors. 

In conclusion, we have shown that DCC formation can explain any excess or
deviation from expectation of hyperons and antihyperons yields based on 
statistical models. Of course there is more than one way to do this kind of
fits; see, for example \cite{bm}. They may or may not accommodate our 
interpretation presented here. However, both the theory and the phenomenology 
based on measured data seem to be self-consistent. For further details one can 
consult \cite{kw,kw2}.

\end{document}